# Survey of Insurance Fraud Detection Using Data Mining Techniques

H.Lookman Sithic, T.Balasubramanian

*Abstract*— With an increase in financial accounting fraud in the current economic scenario experienced, financial accounting fraud detection has become an emerging topics of great importance for academics, research and industries. Financial fraud is a deliberate act that is contrary to law, rule or policy with intent to obtain unauthorized financial benefit and intentional misstatements or omission of amounts by deceiving users of financial statements, especially investors and creditors. Data mining techniques are providing great aid in financial accounting fraud detection, since dealing with the large data volumes and complexities of financial data are big challenges for forensic accounting. Financial fraud can be classified into four: bank fraud, insurance fraud, securities and commodities fraud. Fraud is nothing but wrongful or criminal trick planned to result in financial or personal gains. This paper describes the more details on insurance sector related frauds and related solutions. In finance, insurance sector is doing important role and also it is unavoidable sector of every human being.

## I. INTRODUCTION

Insurance is a contract (policy) in which an individual or entity receives financial protection or reimbursement against losses from an insurance company. The company pools clients' risks to make payments more affordable for the insured. Now a days, customer choosing insurance policy for three reason. (i) risk factor (ii) investment (iii) tax. Today, human life is getting some unexpected movement in their life such as accident, health problem. The accident will make a huge loss for human being. So every one of our people want to save their own life and they need a back up to serve their life. The insurance company helping a people who met accident and suffering from health problem.

But, at the same time now a days the insurance sector facing more problem due to two reason (i) internal side and (ii) external side for eg. Customer claim frauds in insurance sector and lack of knowledge of internal auditor. Now a days, customer entering all the particulars through paper based documents. Then the document scanned and converted as pdf format[9]. After that, the data entered into data base manually by the internal auditor. Due to carelessness and lack of knowledge of internal auditor some misrepresentation will occur for eg. Date of birth entered after the death date. This things will make unexpected loss for insurance sector and also for customer. Next, Every customer should surrender all the legal documents at the time of joining in any insurance sector and also provide right document to collect amount from insurance sector. It is not happening at the time of insurance policy processing.

Due to that two reason some loss occur in insurance sector This paper will focus some common fraud in insurance and detection techniques.

### A. Insurance generally classified into Four types

(i) home insurance

(ii) life insurance

(iii) motor insurance

(iv) medical insurance.

Among four the motor insurance and medical insurance sector has more fraud problems. This paper is describing the details of motor and medical insurance fraud detection.

### B. Various stages of insurance fraud

Fraud can occur at any stage during the process of applying, using, buying, selling, underwriting insurance or while staking a claim which can be broadly categorized as pre-insurance otherwise known as application fraud and post insurance comprising eligibility and claims fraud.

*Pre insurance stage: Application Fraud:*

Application fraud has committed when material misrepresentations are made on an application for insurance with the intent to defraud. Planned non-disclosure by clients has always been a major problem faced by insurance industry, which sadly is socially acceptable.

### i. Post insurance stage: claims fraud:

Most common where the losses are not real, overstated, manipulated, to name a few. Size and occurrence of any insurance fraud is greater at the claim stage in comparison to pre – insurance stage[1].

### C. Types of insurance claims frauds

### i. Motor insurance frauds:

Motor insurance is the most possible and weak fraud ridden sector in the sector in comparison to other line of insurance. motor own damage claims fraud committed at pre and post insurance stage . Auto mobile insurance data are usually binary indicator grouped into accident, claimant, driver, injury, treatment, lost wages, vehicle, and other categories.

There are no publicly available data sets for studying fraud detection except for a relatively small automobile insurance data set. And obtaining real data from companies for research purposes are extremely hard due to legal and competitive reasons. To avoid that data availability problems and work on a particular fraud type, one alternative is to create synthetic data which matches closely to actual data.

Manuscript received February, 2013.
**H.Lookman sithic**, Associate.Professor Department of Computer Applications, Periyar University, Muthayammal College of Arts & Science, Namakkal, India.
**T.Balasubramanian**, HOD Cum Asst.Professor, Department of Computer Sciences, Periyar University/ Sri Vidya Mandir Arts and Science College/ Krishnagiri, India.

# Survey of Insurance Fraud Detection Using Data Mining Techniques

*Generally the motor fraud has two types:*

**i. Hard frauds:**

it includes total damage to the vehicle deliberately to get rid of the same or to earn money than its market value. Some of the examples are theft the vehicle, burnt by fire, fall into river, loss under an excluded risk etc. A real accident may occur, but the dishonest owner may take the opportunity to incorporate a whole range of previous minor damage to the vehicle into the garage bill associated with the real accident.

**j. Soft fraud:**

It accounts for the majority of the motor insurance frauds. Eg: more than one claim for single loss, higher cost for repair, damage caused earlier, replacement of old spare parts etc.

### D. Health Insurance

Health insurance systems are either sponsored by governments or managed by the private sector, to share the health care costs in those countries. Health insurance fraud is described as an intentional act of deceiving, concealing or misrepresenting information that results in health care benefits being paid to an individual or group.

"According to the National Health Care Anti-Fraud Association, health care fraud is an intentional deception or misrepresentation made by a person or an entity that could result in some unauthorized benefit to him or his accomplices".

The medical insurance fraud characteristics include: Damage level insufficient information, suspected diagnosis of proof, insured low willingness to cooperate and cause of the accident unreasonable, repeatedly claims record, in a special area, occur at a specific time and claims for late filing. Inconsistent documents of application, high claim payments, certificate of poor reliability, non –cooperation and very familiar with insurance knowledge.

Each claim is submitted by an affiliate under the approval of a medical professional justifying the work incapacity. Data such as age, sex, type of claim, affiliate's name and date of birth, ID number, resting period solicited, type and place of the resting, identification of the medical professional, identification of the employer, labor activity of the company where the affiliate works, affiliate's profession and income records that the affiliate has gotten in the last three months are incorporated in each form.

A neural classifier that makes a predictive detection of the fraudulent and abusive claims would be of great help for the medical experts in their reviewing process, acting as a pre screen filter. This predictive detection must only consider historic data associated to the affiliate, the medical professional and the employer, and data available before the medical revision of the arriving medical claim.

If the data are consistent, check by manual comparison whether the hospitals and doctors issuing the certificates are registered medical institutions and doctors. Then, by comparison with historical cases and records of the insurant, check whether the case is questionable. As the health insurance is national wide so it takes 45 days minimum to complete the process. So, the research is to solve manpower and to take right decision making by utilization of data mining technology.

False claims are the most common type of health insurance fraud. The goal is to obtain unmerited payment for a claim or series of claims. Health industry in india is loosing approximately Rs 600 crore on "false claims" every year. So to make health insurance feasible there is a need to focus on eliminating or reducing fraudulent claims.

*Hard fraud*: it is a deliberate attempt either to point an event or an accident, which requires hospitalization or other type of loss that would be covered under a medical insurance policy.

*Soft fraud*: it may also occur when people purposely provide false information in regard to the pre-existing illness or other relevant information to influence the underwriting process in the favor of the applicant. Such as, claim fraud, application fraud, eligibility fraud.

### E. Fraud claims trigger

1. Treatment costs are usually on the higher side as compared to the etiology.
2. Patients residence and the hospital address are not geographically same.
3. Higher per patient cost
4. Higher per patient test
5. Fluctuating monthly claims of the providers
6. Double billing.

## II. DATA MINING

Data mining is to find out information with special meaning from a great number of data by some technology as the procedure to discover knowledge from the database, the steps are as follows[11]:

(a) Data cleaning: remove noise data and inconsistent data.
(b) Data integration: combine various data source.
(c) Data selection: find out data related to the subject from the database.
(d) Data Transformation: transform data into the form appropriate for mining.
(e) Data Mining: extract data models by the utilization of technology.
(f) Pattern evaluation: evaluate models really useful to present knowledge.
(g) Knowledge presentation: present knowledge after mining to the users by using technology such as visual presentation.

So, the data mining is to adopt the mining procedure to discover unknown knowledge and rules from plentiful data. Data mining which is part of an iterative process called knowledge discovery in database can assist to extract this knowledge automatically. it has followed better direction and use of health care fraud detection and investigative resource by recognizing and quantifying the underlying attributes of fraudulent claims, fraudulent providers and fraudulent beneficiaries.

### A. Process Model of Data mining in financial fraud detection

Nowadays, auditors are using data mining tools and techniques to examine the entire population of transactions in order to select samples for test controls and identify fraud. The process model for data mining can be successfully used for fraud detection data mining projects[10]. These can be shown as below

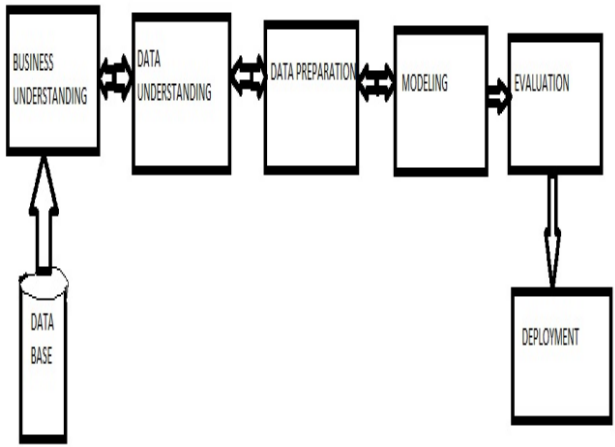

Figure 1: Data mining Techniques for insurance fraud detection

*B. Classification*

Classification builds up and utilize a model to predict the categorical labels of unknown objects to distinguish between objects of different classes. These categorical labels are predefined, discrete and unordered[7]. classification is the process of identifying a set of common features and proposing models that describe and distinguish data classes or concepts[4].

Common classification techniques are neural networks, the naïve Bayes technique, decision trees and support vector machines. Such classification tasks are used in the detection of credit card, health care and automobile insurance, and corporate fraud.

*C. Clustering*

Clustering is used to partition objects into previously unknown conceptually meaningful groups. with the objects in a cluster being similar to one another but very dissimilar to the objects in other clusters. Clustering is also known as data segmentation or partitioning and is regarded as a variant of unsupervised classification [7].It is an unsupervised data mining technique, deals with the problem of dividing a given set of entities into meaningful subsets. Claims with similar characteristics are grouped together and small population clusters are flagged for further investigation for eg. In group insurance, Large beneficiary payment, large interest payment amounts and long lag between submission and payment.

The most common clustering techniques are the K-nearest neighbor, the Naïve Bayes technique and self organizing maps and K means.

*D. Prediction*

Prediction estimates numeric and ordered future values based on the patterns of data set[5]. It is noted that, for prediction, the attribute,for which the value being predicted is continuous –valued rather than categorical. This attribute is referred as the predicted attribute[7]. Neural networks and logistic model prediction are the most commonly used prediction techniques.

*E. Outlier detection*

Outlier detection is employed to measure the "distance" between data objects to detect those objects that are grossly different from or inconsistent with the remaining data set[7]. Data that appear to have different characteristic than the rest of the population are called outliers. A commonly used technique in outlier detection is the discounting learning algorithm[6].

*F. Regression*

Regression is a statistical methodology used to reveal the relationship between one or more independent variables and a dependent variable that is continuous valued[7]. Many empirical studies have used logistic regression as a benchmark. The regression technique is typically undertaken using such mathematical methods as logistic regression and linear regression, and it is used in the detection of credit card, crop and automobile insurance and corporate fraud.

*G. Visualization*

Visualization refers to the easily understandable presentation of data and to methodology that converts complicated data characteristics into clear patterns to allow users to view the complex patterns or relationships uncovered in the data mining process[3]. The researchers have exploited the pattern detection capabilities of the human visual system by building a suite of tools and applications that flexibly encode data using colour, position, size and other visual characteristics. Visualization is best used to deliver complex patterns through the clear presentation of data or function[8].

III. CONCLUSION

This survey paper categories and summaries from almost all published technical and review articles in insurance fraud detection . it defines the professional fraudster, formalizes the main types and sub types of known fraud. Data mining is growing in the insurance field and more and more each day. As data mining will continue to better the insurance sector. Data mining organizes the data in such a way that it makes the task easier for insurance to auditing and also data mining used to prevent the false claim. In india , the insurance statistics is alarming. According to a survey conducted by one of the leading TPAs, the estimated number of false claims in the industry is estimated at around 10-15 percent of total claims.The report suggest that the insurance industry in india is losing approximately Rs.600 crore on false claims every year. We may not eliminate fraud but we can surely reduce it. Currently we are using artificial data to detect insurance data .so, The future work is to use real data for insurance fraud detection.

## AUTHOR'S PROFILE

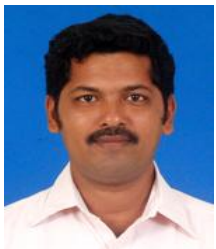

**H.Lookman Sithic**, Received B.Sc Computer Science From Jamal Mohamed College, Bharathidasan University, Trichy 1999, M.S Information Technology From Jamal Mohamed College, Bharathidasan University, Trichy 2001, M.Phil Computer Science from Periyar University, Salem 2004 & Now he is pursuing his Ph.D Computer Science in Bharathiyar University from 2011. He is having 11 Years of Teaching experience in Computer Science, currently he is working as an Associate.Professor Department of Computer Application, Muthayammal College of Arts & Science, Periyar University.

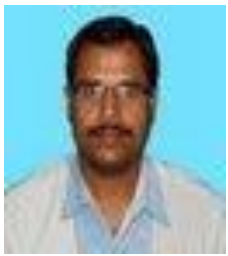

**T.Balasubramanian**, Received B.Sc Computer Science From Bishop heber College, Bharathidasan University, Trichy 1995, M.Sc Computer Science From Jamal Mohamed College, Bharathidasan University, Trichy 1997, M.Phil Computer Science from Periyar University, Salem 2007 & Now he is pursuing his Ph.D Computer Science in Bharathiyar University from 2009. Currently he is working as HOD Cum Asst.Professor Department of Computer Science, Sri Vidya Mandir Arts & Science College.